\documentclass[aps,pre,preprint,superscriptaddress,showpacs]{revtex4}
\usepackage{graphicx}
\begin{document}

\title{Energy Flux Fluctuations in a Finite Volume of Turbulent Flow }

\author{M. M. Bandi}
\affiliation{Department of Physics and Astronomy, University of Pittsburgh, Pittsburgh, PA 15260.}

\author{W. I. Goldburg}
\affiliation{Department of Physics and Astronomy, University of Pittsburgh, Pittsburgh, PA 15260.}

\author{J. R. Cressman Jr.}
\affiliation{Krasnow Institute, George Mason University, Fairfax, VA 22030.}

\author{A. Pumir}
\affiliation{Institut Non Lineaire de Nice, Centre National de la Recheche Scientifique and Universite de Nice, 1361 Route des Lucioles, Valbonne, France 06270.}


\date{\today}

\begin{abstract}
The flux  of turbulent kinetic energy from large to small spatial scales is measured in a small domain $B$ of varying size $R$.  The pdf of the flux is obtained using a time-local version of Kolmogorov four-fifths law.  The measurements, made at moderate Reynolds number, show frequent events where the flux is backscattered from small to large scales, their frequency increasing as $R$ is decreased.  The observations are corroborated by  a numerical simulation based on the motion of many particles and on an explicit form of the eddy damping.
\end{abstract}


\maketitle


\section{Introduction}
 There is considerable interest in the global (spatially averaged) fluctuations in systems driven far from equilibrium \cite{cohen1997,gallavotti1998,pinton1998,aumaitre2001}, of which fluid turbulence provides a striking example \cite{kolmogorov1941,frisch95}.  An essential aspect of three-dimensional turbulence is the cascade of energy from large to small scales, followed by dissipation at the smallest scales. Characterizing the energy flux is particularly important for turbulence modeling. The local rate of energy dissipation is known to fluctuate wildly \cite{frisch95}.  This work investigates flux of energy from large to small scales, averaged over a local region of finite extent.

The spatially averaged value of the energy  flux  $\varepsilon$  over a subvolume of the fluid $B$ of typical size $R$, is  simply given by the rate of dissipated kinetic energy if the system is in a steady state.  In that case the  flux is necessarily positive and is directed from large to small scales. However, temporal fluctuations in this flux can be very substantial. In fact, it is has been demonstrated several times that energy may backscatter from small to large scales, leading to a {\it negative} value of the energy flux \cite{Tao2002,Tsinober2005}. It is natural to expect that this effect should depend on the scale of the subsystem investigated.  One of the aims of this work is to quantify the fluctuations of the energy flux measured over a subdomain of the flow, and in particular, its dependence on the subdomain's size. 

Under conditions of local isotropy, the ensemble averaged energy dissipation rate, $\overline{\varepsilon}$, is related to the third moment of the longitudinal velocity differences at a given scale, $r$ :

\begin{equation}
\label{kolmogorov}
\langle (\Delta u_{L}(r))^{3}\rangle = -\frac{4}{5}\overline{\varepsilon}r
\end{equation}

 The recent theoretical \cite{Tanveer99,Robert,Eyink2003} and numerical work \cite{kurien2003} leads to a generalisation of Eq.~\ref{kolmogorov}, which is local in space and time. More precisely, if one considers any finite subdomain of the flow, $B$, the average of $(\Delta u_L(r))^{3}$ over all directions of the vector $\bf r$, and over all points in $B$, is equal to ${ - 4 \over 5 }\varepsilon_B(t) r $, where $\varepsilon_B$ is the energy flux towards small scales in the subdomain $B$ :

\begin{equation}
\label{tkolmogorov}
\varepsilon_{B}(t) \equiv \frac{1}{2} \frac{d \langle v^2 \rangle}{dt} 
=  \frac{\langle(\Delta u_L(r,t))^{3}\rangle_{B,\Omega}}{-\frac{4}{5}r}
\end{equation}

Here $\langle...\rangle_{B,\Omega}$ denotes the average over the subvolume $B$, and over all possible directions ($\Omega$). The derivation of Eq.~\ref{tkolmogorov} is based on a rigorous energy balance for weak solutions of the Navier-Stokes equations, in the $Re \rightarrow \infty$ limit \cite{Robert}. In this sense the quantity $\varepsilon_B(t)$ is really the instantaneous, inertial range dissipation rate, interpreted here as an energy flux (defined as positive towards small scales). Eq.~\ref{tkolmogorov} justifies the intuitive estimates of the rate of change of energy in a volume, by a straightforward averaging with the help of Eq.~\ref{kolmogorov}, even when the key assumption of local isotropy is not satisfied. The spherical average in Eq.~\ref{tkolmogorov} permits the recovery of the isotropic sector from an arbitrary (anisotropic) flow \cite{kurien2003}. 

In practice, the average over all directions is replaced here by an average over particle pairs along all azimuthal directions in an instantaneous snapshot within the subdomain $B$ of interest. The magnitude and sign of the numerator on the RHS fluctuates in time, though it must be negative on the average if the system is in the steady state. This assures that the time average of the flux $\varepsilon_{B}(t)$ is positive.  

  Averaging over all pairs of particles in a sub-domain of the fluid, as done here experimentally, has also been proposed as a way of estimating the energy dissipation in a numerical scheme, based on the ``smooth particle hydrodynamics'' (SPH) method \cite{monaghan1992}. In such an approach the values of various hydrodynamic fields at a point $x$ are obtained by interpolating, with the help of all the neighbouring particles, within a smoothing distance $h$ from the point $x$. An expression for the eddy-damping, very similar to Eq.~\ref{tkolmogorov} has been postulated in \cite{pumir2003} (see also discussion below). Implementing numerically this eddy-damping in a SPH code, as done in \cite{pumir2003} leads to a Large Eddy Simulation scheme, based on particles moving with the flow (lagrangian particles), where the smoothing length $h$ is the smallest scale resolved numerically. The present experimental setup allows an explicit measurement of the energy flux from Eq. ~\ref{tkolmogorov} and a comparison with the postulated form of the eddy-damping \cite{pumir2003}.  The experimental results demonstrate that the postulated form of the energy-dissipation term is very strongly correlated with the energy flux obtained from Eq.~\ref{tkolmogorov}. Furthermore, the numerical scheme based on SPH allows one to study the energy flux in a subdomain of the system as a function of size. The qualitative agreement between the experimental and the numerical studies is excellent.

\section{Experimental Method}
\begin{figure}
\includegraphics{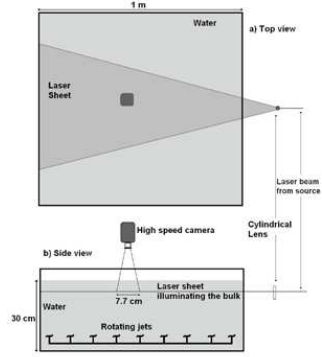}
\caption{\label{expsetup} a) Top view and b) Side view of the experimental setup. A tank of lateral dimensions 1 m x 1 m is filled with water to a depth of 30 cm. A laser beam passing through a cylindrical lens generates a sheet of laser that illuminates a horizontal plane of the turbulent fluid 4 cm below the surface. Turbulence is generated by a system of 36 capped rotating jets situated at the tank floor. Neutrally buoyant tracer particles are suspended in the turbulent fluid. A high speed camera suspended above the tank records the motion of tracers as they scatter light upon entering the sheet of illumination.}
\end{figure}

 The experimental arrangement is identical to the one used previously for studying surface turbulence \cite{RobNJP2004,RobThesis} and is shown in Fig. \ref{expsetup}. Measurements are made 4 cm below the surface in a square tank of side length 1 m, filled with water to a height of 30 cm. The fluid is seeded with neutrally buoyant glass beads of 10 $\mu$m mean diameter. The local velocity of the fluid is captured by a high speed camera that records the movement of particles over a square area of side length L = 7.7 cm. The laser sheet illuminating the particles is roughly 1 mm thick, but the camera's depth of field is a tenth of this. Thus only the horizontal components of velocity are captured. The buffer size of the camera limits the duration of data collection to 5 s at a frame rate of 400 frames per second. The velocity field is constructed by correlating pairs of images with a particle tracking program. The spatial resolution of the measurements depends upon the length of a pixel in the camera's sensor. Hence the spatial resolution in this experiment is 7.7 cm/1024 pixels = 75 $\mu$m. The temporal resolution is the sampling rate which is 7.5 ms for this experiment. Hence the velocity resolution obtained in this experiment is 1 cm/s. To ensure reliability of the measured velocities, every $n^{th}$ and $(n+3)^{rd}$ image is employed in velocity field construction. The weak non-zero mean velocity observed in the flow is systematically subtracted out. Physical parameters that characterize the turbulence are given in table~\ref{tab1}. 

\begin{table}
\caption{\label{tab1}Turbulent quantities of interest measured in the experiment.}
\begin{ruledtabular}
\begin{tabular}{|c|c|c|}
Parameter  & Expression  & Measured Value\\ \hline
Taylor Microscale $\lambda$ (cm) & $\sqrt{\frac{u_{rms}^{2}}{\langle(\partial u_{x}/\partial x)^{2}\rangle}}$ & 0.32\\
Taylor Microscale Reynolds Number $Re_{\lambda}$  & $\frac{u_{rms}\lambda}{\nu}$  & 79\\
Integral Scale $l_{0}$ (cm) & $\int dx (\frac{\langle(u_{L}(x+r)u_{L}(x))\rangle}{\langle u_{L}(x)^{2}\rangle})$ & 3.5\\
Large Eddy Turnover Time $\tau_{l_{0}}$ (s) & $\frac{l_{0}}{u_{rms}}$ & 1.4\\
Dissipation Rate $\varepsilon_{diss}$ $(cm^{2}/s^{3})$  & $15\nu\langle (\frac{\partial u_{x}}{\partial x})^{2}\rangle$  & 8.3\\
Kolmogorov Scale $\eta$ (cm) & $(\frac{\nu^3}{\varepsilon_{diss}})^{1/4}$ & 0.02\\
RMS Velocity $u_{rms}$ (cm/s) &  $\sqrt{\langle u^{2}\rangle - \langle u \rangle^{2}}$ & 2.4\\
\end{tabular}
\end{ruledtabular}
\end{table}

\section{Results and Discussion}
\begin{figure}
\includegraphics{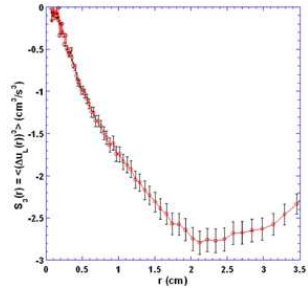}
\caption{\label{s3_timeavg}(Color online) The time averaged third-moment of longitudinal velocity differences ($S_{3}(r)$) as a function of particle separation r (cm). The plot is constructed by averaging over twelve time-uncorrelated velocity snapshots.}
\end{figure}

  Though the Reynolds number in the experiment is low, the flow exhibits a well-defined inertial range. Figure~\ref{s3_timeavg} shows the space and time-averaged $S_3(r) =  \langle \Delta (u_L(r))^3 \rangle $ vs. $r$ in the range 0.07 cm $\leq r \leq$ 3.5 cm. Each instantaneous third-moment is constructed from an average over all particle pairs in the field of view. There are roughly $10^{5}$ particle pairs at each r-value contributing to the spatial average at each instant. Twelve time-uncorrelated measurements of the third-moment are averaged to obtain the plot in Fig.~\ref{s3_timeavg}. The interval between these measurements is 1.4 s, which is the lifetime of the largest eddies, as determined from the velocity autocorrelation function. Therefore each data-set provides 4 measurements uncorrelated in time. Three such data-sets are taken in quick succession to obtain the average. In forming this spatial average of $S_{3}(r)$, data for $r < 0.07$ cm are excluded, since velocity differences cannot be reliably measured below this value. The statistical error in the time averaged $S_3(r)$ is less than 5$\%$ as observed from the error bars in fig. \ref{s3_timeavg}.

\begin{figure}
\includegraphics{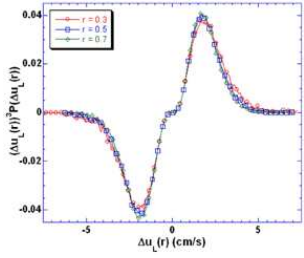}
\caption{\label{s3integrand}(Color online) The integrand of the third-moment of longitudinal velocity differences can be constructed reliably for an instantaneous snapshot for different spatial separations $r$ (here taken for $r$= 0.3, 0.5 and 0.7 cm). }
\end{figure}

  Shown in Fig.~\ref{s3integrand} is $x^{3}P(x)$ where $x= \Delta u_{L}(r,t)$ at the three indicated $r$-values 0.3, 0.5 cm, and 0.7 cm.   The measurements were made at one instant of time.  The integral of this function over $x$ is the the instantaneous third moment of interest.  Each $P(\Delta u_{L}(r,t))$ is constructed from $10^5$ particle pairs.  The statistical fluctuations  in each of the curves is adequately small to yield a statistically significant value of the (fluctuating) third moment and hence the pdf of this moment $P(S_{3}(r))$, at least for the values of $r$ listed  above.  According to Eq. ~\ref{tkolmogorov}, the pdf of the energy flux is then given by $P( \varepsilon)= P(S_3(r))/(-\frac {4}{5}r)$.  

\begin{figure}
\includegraphics{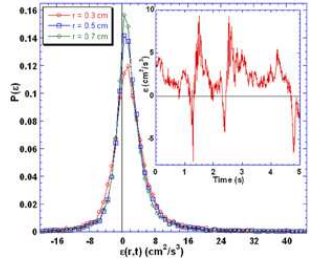}
\caption{\label{exp_flux_pdf}(Color online) The inset shows the time trace of energy flux obtained from Eq.~\ref{tkolmogorov}. The main figure shows the pdf of energy flux obtained for three different spatial separations r = 0.3, 0.5 and 0.7 cm.}
\end{figure}

 The inset to Fig.~\ref{exp_flux_pdf} shows a 5-second time record of $\varepsilon(r,t)$ for $r$ = 1.925 cm.  One sees that the energy flux is positive most of the time, but does show intervals of reversed energy flow. The figure itself shows the pdf of energy flux at the same three inertial range $r$-values of Fig.~\ref{s3integrand}.  One expects that in the inertial range all three curves should coincide, and they approximately do so. This $r$-independence in the inertial range is expected for the third-moment. The energy flux here is experimentally obtained from Eq.~\ref{tkolmogorov} for a sub-domain $B$ of size $R = 1.925$ cm. As required by Eq.~\ref{tkolmogorov} the pdfs are positively skewed.

  The energy flux is also experimentally obtained for sub-domains $B$ of size $R$ ($R$ = 7.7 cm, 3.85 cm and 1.925 cm) keeping the particle separation $r = 0.7$ cm constant.  The ``width'' of their pdfs decreases with increasing $R$, as one would expect.  In general agreement with previous observations, it is found  that the probability of backscattering is significant, but it is measured to be a monotonically decreasing function of $R$ when $R$ increases beyond $l_0$. 

\section{Numerical Simulations}
 The experimental results have been supplemented by a numerical simulation, using the SPH algorithm to simulate flows with particles \cite{monaghan1992}. In this approach, the fluid is represented by particles, whose positions, ${\bf r}_i$  and velocities, ${\bf v}_i$ evolve according to the equation of motion :
\begin{equation}
{ d {\bf r }_i \over dt } = {\bf v}_i ~~~; {d {\bf v}_i \over dt } = - \nabla p_i+ {\bf f}_i + {\bf D}_i  
\label{eq_motion_SPH}
\end{equation}
where $\nabla p_i$ is the pressure gradient, ${\bf f}_i$ the forcing term, and ${\bf D}_i$ the dissipation term. The pressure is obtained from the local density $\rho_i$ from the equation of state $p = \rho^{\gamma}$ ($\gamma=9$), which corresponds to a very weakly compressible gas provided $|v_i|$ is small compared to the velocity of sound. The density $\rho$ as well as the gradients are computed at a point $\bf x$ by interpolating the properties of the particles with a kernel function characterized by a size $h$. In this approach, scales smaller than $h$ are unresolved. The SPH method allows one to simulate flows with a resolution no better than $h$. 

  The issue is thus to adequately parametrize the sub-grid energy dissipation $\overline\varepsilon_{h}$ occuring below the scale $h$. The precise form for $\overline\varepsilon_{h}$ is inspired by the Kolmogorov equation \cite{pumir2001,pumir2003};  it corresponds to an energy dissipation of :
\begin{equation}
\overline\varepsilon_{h} = -\frac{\nu_{t}}{h^{2}}\langle({\bf v}_{ij}.{\bf r}_{ij}) {\bf v}_{ij}^{2} \rangle.
\label{dissip_SPH}
\end{equation}

\begin{figure}
\includegraphics{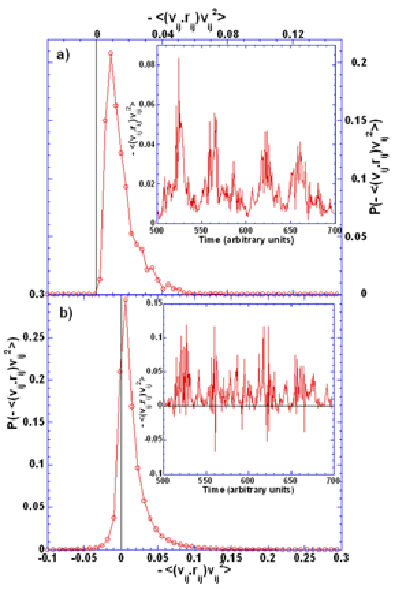}
\caption{\label{sim_flux_pdf}(Color online) (a) shows the time-trace (inset) and the pdf of energy flux for a simulation box of side length L/2. The fluctuations are always positive and never change sign. (b) shows the time-trace (inset) and the PDf of energy flux for a simulation box of side length L/8. The fluctuations switch sign frequently when the box size has diminished.}
\end{figure}

Here the angular brackets denote an average over all particles and this term is negative on the average, and ${\bf v}_{ij} \equiv ({\bf v}_i - {\bf v}_j)$ is the velocity difference between the particles $i$ and $j$ separated by the distance $r_{ij}$. The parameter $\nu_{t}$ is a dimensionless constant deduced to be of the order of unity \cite{pumir2003}, and should not be confused with kinematic viscosity.  The experimental apparatus permits  measurement of the postulated eddy damping term, and a  comparison with the form of Eq.~\ref{tkolmogorov}. Agreement between the flux in Eq.~\ref{tkolmogorov} and the dissipation in Eq.~\ref{dissip_SPH} is extremely good. With a properly adjusted value of $\nu_t$, the correlation between $\varepsilon(t)$ and $\varepsilon_{h}$ is $\sim 0.98$ at the three values of $R = 7.7$, $3.85$ and $1.925$ cm. All the properties of the energy flux computed with Eq.~\ref{dissip_SPH} are essentially identical to the properties computed with Eq.~\ref{tkolmogorov}, as shown in Fig. ~\ref{sim_flux_pdf} . The precise value of $\nu_t$ is $\sim 4.8$. The value of $\nu_t$ depends on the precise geometry of the subdomain $B$. This experimental finding is a striking confirmation of the validity of the eddy-damping postulated in \cite{pumir2003}.

  The implementation of SPH algorithm is described in \cite{pumir2003}.In the calculation, the forcing term is chosen as a random superposition of three low-wavenumber fourier modes. The velocity of each particle was no larger than $1/10$ the velocity of sound, so the flow is effectively incompressible.  The energy flux over a sub-volume of the system is computed numerically, as effectively done in the experiment. The total energy flux averaged over a system volume of side-length L/2 (for a simulation box of side-length L) has large fluctuations. The inset to Fig.~\ref{sim_flux_pdf}(a) shows that the fluctuations remain negative for as long as the simulations have been carried out. When averaged over a smaller fraction of the volume of side-length L/8, one finds that positive fluctuations of the averaged energy-flux become possible. This is evident from the inset to Fig.~\ref{sim_flux_pdf}(b) and the corresponding pdf. The smaller the box size larger are the fluctuations. This finding is fully consistent with the experimental observations. As in the experiment, it was found that the probability of energy flux backscattering is a monotonically decreasing function of the scale $R$ over which the flux is defined.

\section{Summary}
Fluid turbulence research has focussed either on small scale properties, with the objective of understanding ``intermittency'', or on averages of global properties of interest for most applications. Only recently has a more systematic investigation of global fluctuations been undertaken \cite{pinton1999, pinton1998, cadot2004}.  The focus here is on the energy transfer to smaller scales, a hallmark of hydrodynamic turbulence. Whereas on average, the fluid transfers energy to smaller scales via the cascade process, the observed fluctuations are so large, as to actually reverse the energy flux from small to large scales \cite{Tao2002}. This experimental study of the fluctuations of the energy flux is based on an experiment carried out at moderate Reynolds number. The results are based on Eq.~\ref{tkolmogorov}. It has been shown that the probability of energy backscattering decreases when the size of the system increases. Also, the form of the eddy-damping term as proposed in \cite{pumir2003} has been justified. The experimental results are corroborated by the results of the calculation based on the particle method.

 The systematic study of scale dependence of the fluid properties is a novel aspect of this work. It would be interesting to understand these results in the light of recent fluctuation theory derived in the context of nonlinear, out-of-equilibrium systems \cite{evanscohen1993, gallavotti1998}.


%


\begin{acknowledgments}
MMB and WIG acknowledge S. Bhattacharya for help with initial measurements and helpful discussions with the University of Pittsburgh Softmatter Group, B. Eckhardt, I. Procaccia and S. Kurien. AP acknowledges helpful conversations with J.-F. Pinton and G. L. Eyink. This work was supported by the NSF (Grant No. DMR-0201805). AP was supported by the grant HPRN-CT-2002-00300 from the European Commission, and by IDRIS for computational resources.
\end{acknowledgments}

\bibliography{ff122305}

\end{document}